\documentclass[twocolumn,showpacs,preprintnumbers,amsmath,amssymb,floatfix]{revtex4}

\usepackage{epsfig,psfrag}
\usepackage{dcolumn}
\usepackage{bm}
\usepackage{graphicx}
\usepackage{color}

\begin{document}

\title{Magnetic confinement of massless Dirac fermions in graphene}
\author{A.~De~Martino, L.~Dell'Anna, and R. Egger}
\affiliation{Institut f\"ur Theoretische Physik, Heinrich-Heine-Universit\"at,
D-40225  D\"usseldorf, Germany}
\date{\today}

\begin{abstract}
Due to Klein tunneling, electrostatic potentials are unable to confine
Dirac electrons. We show that it is possible to confine  massless
Dirac fermions in a monolayer graphene sheet by inhomogeneous magnetic fields.
This allows one to design mesoscopic structures 
in graphene by magnetic barriers, e.g. quantum dots or quantum
point contacts.
\end{abstract}
\pacs{73.21.-b, 73.63.-b, 75.70.Ak}

\maketitle

The successful preparation of monolayer graphene films \cite{geim,kim,walt}
has recently generated a lot of excitement and allows one to directly probe
the physics of two-dimensional (2D) Dirac-Weyl fermions. 
The massless Dirac spectrum at low energy scales
is caused by the sublattice structure (the basis of graphene's
honeycomb lattice  contains two carbon atoms, giving rise
to an isospin degree of freedom) combined with a 
special band structure, and has been verified experimentally
\cite{geim,kim,zhou}.  Besides the fundamental interest, 
graphene has also been suggested as a building
block for future nanoelectronic devices \cite{walt}.  However, there
is an interesting twist at that point, since
Dirac fermions cannot immediately be confined by electrostatic potentials.
In marked contrast to the Schr\"odinger case,  
Dirac fermions can penetrate high and wide electrostatic barriers with 
high transmission probability, in particular for normal incidence.
This is often referred to as {\sl Klein tunneling} \cite{calo}
and can be understood by noting that under the barrier, the whole
spectrum is shifted upwards.  Incoming electron-like
quasiparticles can then efficiently tunnel through the barrier
via empty states in the hole band, which are always available
since the Dirac spectrum is unbounded.
In the context of carbon nanotubes and graphene, this effect was 
theoretically studied in Refs.~\cite{ando,falko,geim2}.
The creation of useful mesoscopic structures, 
e.g.~quantum dots or quantum point contacts,
thus seems to encounter a major and fundamental 
obstacle, seriously limiting graphene's potential for applications.

An obvious but rather crude way out of this dilemma
is to mechanically cut samples into the desired shape.
Alternatively, one could attempt to exploit the fact that
suitable transverse states in a graphene strip may allow one
to circumvent Klein tunneling \cite{efetov}.  Here we describe a 
completely different and hitherto unnoticed way of confining
Dirac-Weyl quasiparticles in graphene by {\sl magnetic barriers}. 
Employing existing technology, the required {\sl inhomogeneous} 
static magnetic field configurations can be created using 
ferromagnetic layers located beneath the substrate on which the graphene layer
is deposited; for other possibilities, see Ref.~\cite{lee}.
Mesoscopic transport with magnetic barriers has
been experimentally studied for the Schr\"odinger fermions realized  
in conventional semiconductor heterostructures, e.g.~transport 
in the presence of magnetic barriers \cite{expbar} and
superlattices \cite{carmona}, magnetic edge states 
close to a magnetic step \cite{henini}, and magnetically confined 
quantum dots or antidots \cite{antidots}.
Correspondingly, apart from one study of magnetic edge states 
in narrow-gap semiconductors \cite{malkova},
model calculations have only been carried out for
Schr\"odinger fermions \cite{lee,peeters1,peeters2,peeters3},
Here we formulate the theory of magnetic barriers and 
magnetic quantum dots for the massless Dirac-Weyl fermions in graphene.
With minor modifications, the theory also covers narrow-gap semiconductors.

We focus on a static {\sl orbital}\ magnetic field \cite{foot}
oriented perpendicular to the graphene ($x-y$) plane, 
${\bf B}= B(x,y) \hat e_z$, and work on the 
simplest possible theory level (no disorder, no
interactions), where the electronic spin degree of freedom can be disregarded.
Moreover, we consider the physically relevant case of
slow $B(x,y)$ variations on the scale of the graphene
lattice spacing ($a=0.246$~nm).  On low energy scales, 
a weakly doped  (or undoped) graphene layer is described by two 
identical copies of the Dirac Hamiltonian, which remain decoupled in 
the presence of smoothly varying magnetic fields 
\cite{semenoff}.  These two copies describe low-energy envelope states 
in the ${\bf k}\cdot{\bf p}$ approach \cite{mele} close to
the two relevant K points in the hexagonal first Brillouin zone of
graphene.  For slowly varying ${\bf B}={\rm rot} \ {\bf A}$, we therefore 
need to study just one K point. The time-independent Dirac equation for
the spinor $\psi(x,y) = (\psi_+,\psi_-)^T$ 
at energy $E=v_F \epsilon$ then reads (we put $\hbar =1$)
\begin{equation}\label{h0}
\vec \sigma  \cdot  \left({\bf p} +\frac{e}{c} {\bf A}(x,y) \right) \psi(x,y)
= \epsilon \psi(x,y),
\end{equation}
where the Fermi velocity is $v_F\approx 8\times 10^5$~m/sec, 
the momentum operator is ${\bf p}=-i(\partial_x,\partial_y)^T$,
and the $2\times 2$ Pauli matrices in $\vec \sigma=(\sigma_x,\sigma_y)$
act in isospin space. 
The velocity operator follows from the Heisenberg equation as
${\bf v}=v_F \vec \sigma$.
 In this paper, we discuss two prime examples of interest
based on the Dirac-Weyl Hamiltonian in Eq.~(\ref{h0}), namely, 
(i) the magnetic barrier and (ii) a circularly symmetric
magnetic quantum dot.  

For a {\sl magnetic barrier}, the relevant physics is described by a 
magnetic field translationally invariant  along the (say) $y$-direction,
$B(x,y)=B(x)$.  Choosing the vector potential in the
gauge ${\bf A}(x,y)= A(x)\hat e_y$ with $\partial_x A(x)=B(x)$,
transverse momentum $p_y$ is conserved, and for
given $p_y$, Eq.~(\ref{h0}) leads to the coupled equations
\begin{equation} \label{coupled}
[\partial_x \pm p_y \pm (e/c) A(x)] \psi_\pm (x)= i \epsilon \psi_\mp(x).
\end{equation}
These equations imply the decoupled 1D 'Schr\"odinger' equations
\begin{equation} \label{schr}
[\partial_x^2 - V_\pm(x) + \epsilon^2] \psi_\pm (x)= 0,
\end{equation}
with the $p_y$-dependent effective potentials 
\begin{equation}\label{effpot}
V_\pm (x) =  \pm (e/c) \partial_x A(x) + [p_y+ (e/c) A(x)]^2.
\end{equation}
Let us then describe the solution of Eq.~(\ref{schr}) 
for a {\sl square-well magnetic barrier}, where
${\bf B}=B_0 \hat e_z$ (with constant $B_0$) within the strip 
$-d\leq x\leq d$ but ${\bf B}=0$ otherwise, 
\begin{equation}\label{barrier}
B(x,y) = B_0 \ \theta(d^2-x^2) ,
\end{equation}
with the Heaviside step function $\theta$.
The sharp-edge form (\ref{barrier}) is appropriate 
when the Fermi wavelength $\lambda_F$ is parametrically larger than the 
edge smearing length $\lambda_s$,
while $\lambda_s\gg a$ to ensure smoothness of ${\bf B}$;
otherwise scattering between the two K points takes place \cite{foot1}.
Note that here $\lambda_F\sim 1/|\epsilon|$ is determined by the 
(inverse) Fermi momentum of the Dirac quasiparticles,  which is
measured  relative to the relevant K point, and 
the large momentum scale associated with the K point
itself drops out completely.
With the magnetic length $l_B\equiv \sqrt{c/eB_0}$,
the vector potential is written as
\begin{equation}\label{vecpot1}
A(x) = \frac{c}{el_B^2} \times
\left\{ \begin{array}{cc} -d , & x<-d\\   x , & |x|\leq d\\
d, & x>d \end{array} \right. .
\end{equation}
Consider now an electron-like scattering state ($\epsilon> 0$)
entering from the left side, with incoming momentum 
${\bf p}=(p_x,p_y)$. The incoming wave function is, up to 
an overall normalization,
\[
\psi_{in}(x) = \left( \begin{array}{c} 1 \\
\frac{p_x +i (p_y- d/l_B^2)}{|{\bf p}|} \end{array} \right) e^{ip_x x} ,
\]
where the shift in $p_y$ is due to  our gauge choice for the vector potential.
It is then convenient to parametrize the momenta as
\begin{equation}\label{momentumpar}
p_x= \epsilon \cos\phi , \quad p_y = \epsilon \sin\phi + d/l_B^2.
\end{equation}
The gauge-invariant velocity is
${\bf v}= v_F(\cos\phi ,  \sin\phi)^T$, and therefore
$\phi$ is the kinematic incidence angle.
The emergence angle $\phi'$ at the right barrier, 
$p_x'=\epsilon\cos\phi'$, is obtained by exploiting conservation of $p_y$,
\begin{equation}\label{cons}
\sin\phi' =  \frac{2d}{\epsilon l_B^2} + \sin\phi.
\end{equation}
Up to an overall normalization factor,
the scattering state in the three regions is as follows.
For $x<-d$, 
\begin{equation}\label{state1}
\psi_I(x) = \left( \begin{array}{c} 1 \\
e^{i\phi} \end{array} \right) e^{ip_x x}
+r \left( \begin{array}{c} 1 \\ - e^{-i\phi}
 \end{array} \right) e^{-ip_x x}
\end{equation}
with $\phi$-dependent reflection amplitude $r$. 
In the barrier region $|x|\leq d,$ the solution is expressed
in terms of parabolic cylinder functions $D_\nu$ \cite{gradsteyn}, 
\begin{equation}\label{state2}
\psi_{II} = \sum_\pm  c_\pm
\left( \begin{array}{c} 
D^{}_{(\epsilon l_B)^2/2-1} \left (\pm \sqrt{2}(x/l_B+p_y l_B) \right) \\
\pm i \frac{\sqrt{2}}{\epsilon l_B} D^{}_{(\epsilon l_B)^2/2} \left
( \pm \sqrt{2}(x/l_B+p_y l_B) \right ) 
\end{array} \right) 
\end{equation}
with complex coefficients $c_\pm$.
Finally, for $x>d$, the transmitted wave is 
\begin{equation}
\psi_{III}(x) =  t \sqrt{p_x/p_x'} \left(
\begin{array}{c} 1 \\ e^{i\phi'} \end{array}\right) e^{ip'_x x}
\end{equation}
with transmission amplitude $t$.
The transmission probability $T=|t|^2$ is then
related to the reflection probability $R=|r|^2$ by $T+R=1$.
Note that Eq.~(\ref{cons}) implies that 
for certain incidence angles $\phi$,  no transmission is possible.
In fact, under the condition 
\begin{equation}\label{perfecref}
\epsilon l_B \leq  d/l_B ,
\end{equation}
{\sl every}\ incoming state is {\sl reflected}, regardless of the
incidence angle $\phi$. 
In essence, all states with cyclotron radius (defined 
under the magnetic barrier)
less than $d$ will bend and exit backwards again.
This illustrates our main finding: in 
contrast to electrostatic barriers, {\sl magnetic barriers
are able to confine Dirac-Weyl quasiparticles}.  
For sufficiently large barrier width $2d$ and/or field $B_0$, 
all relevant states will be reflected. Our analysis also
shows that this conclusion is generic and  does not depend on the particular
choice (\ref{barrier}) for the barrier.

\begin{figure}[t!]
\scalebox{0.4}{\includegraphics{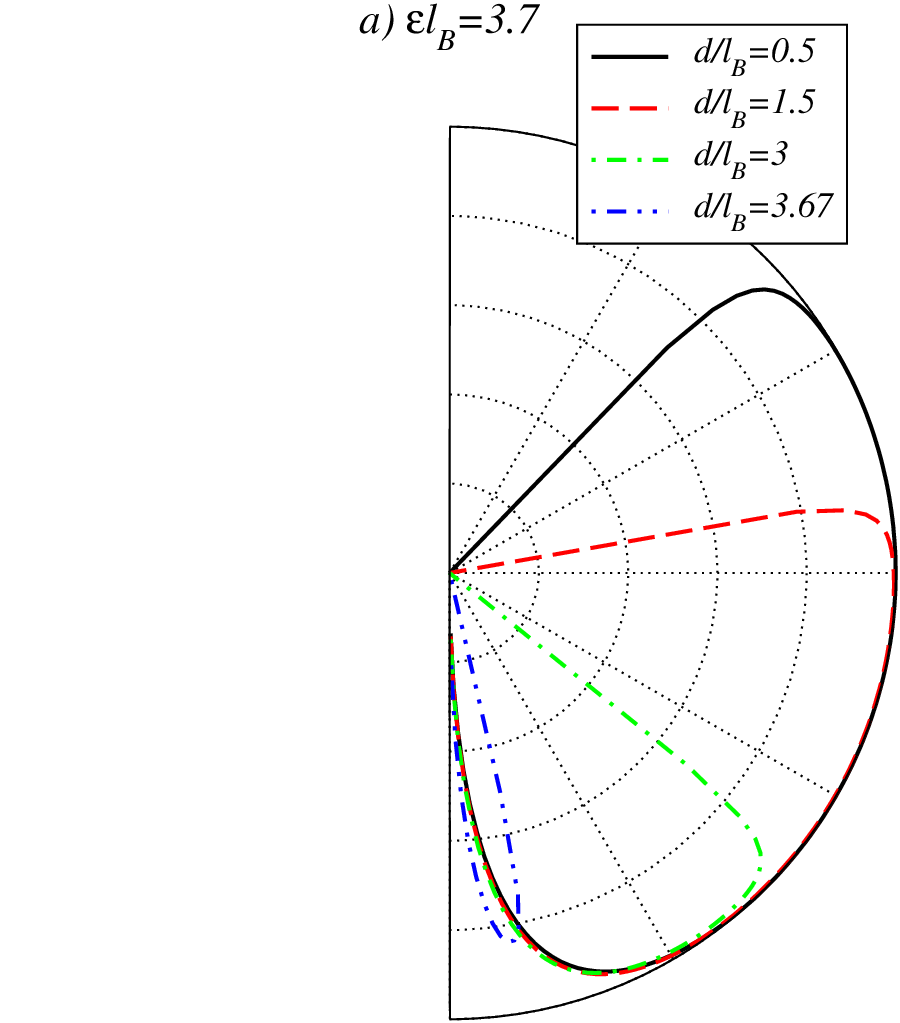}} 

\vspace{1cm}
\scalebox{0.4}{\includegraphics{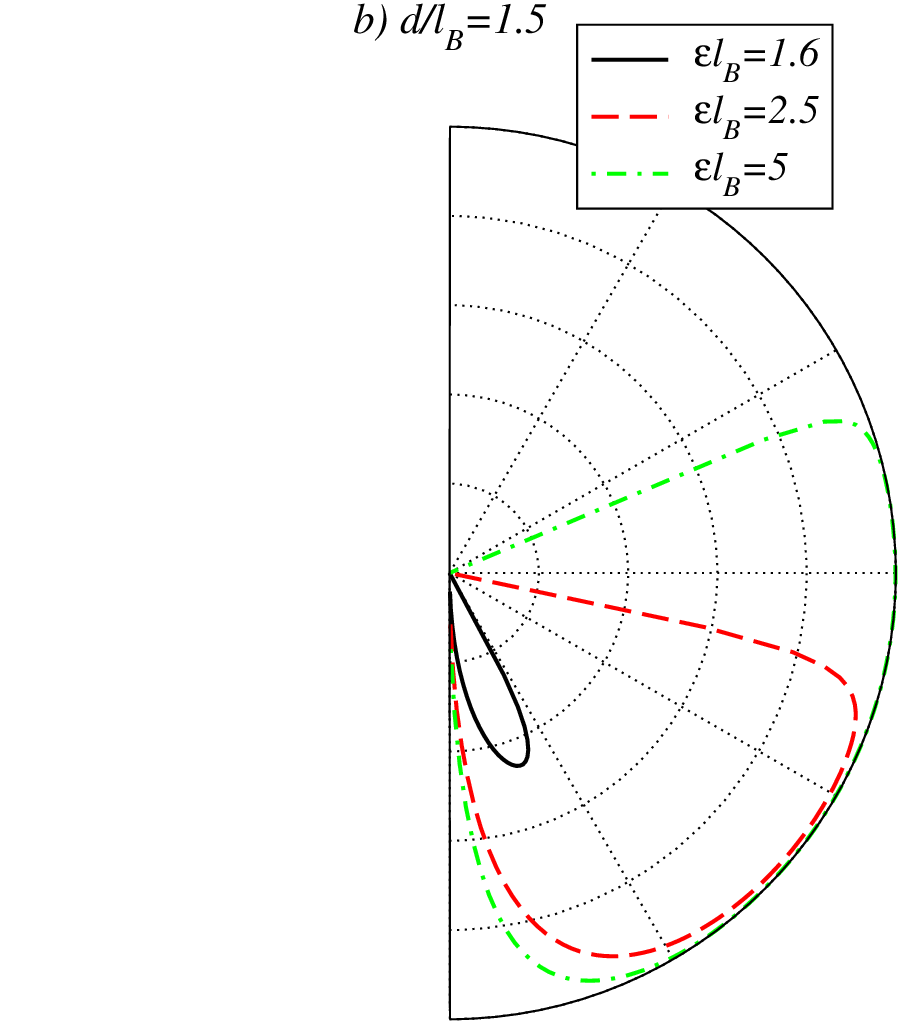}}
\caption{ \label{fig1} (Color online)
Polar graphs depicting the transmission probability $T(\phi)$ 
for a magnetic barrier of width $2d$ at various
energies $\epsilon$.  The outermost semicircle corresponds to $T=1$,
the center to $T=0$, with grid spacing $0.2$.  Angles between
$-\pi/2$ and $+\pi/2$ are shown, the angular grid spacing is $\pi/6$.
(a) $T$ as function of barrier width for fixed energy, $\epsilon l_B=3.7$.
For $d/l_B\geq 3.7$, the transmission is zero for all $\phi$.
(b) Same as function of $\epsilon$ for $d/l_B=1.5$.  The transmission
vanishes for $\epsilon l_B\leq 1.5$.
}
\end{figure}

If the condition (\ref{perfecref}) is not obeyed, 
the transmission probability $T$ does not vanish in general.
Its value follows by enforcing continuity of the wavefunction
at $x=\pm d$.  The solution of the resulting linear algebra problem 
yields the transmission amplitude in closed form,
\begin{eqnarray}\label{trans}
t &=& \frac{2 i \epsilon l_B  \sqrt{2p_x'/p_x} \ 
 \cos\phi}{ e^{i(p_x+p_x')d} \ {\cal D} } ( u_2^+ v_2^- + v_2^+ u_2^- ),
\\ \nonumber
{\cal D} &=& (\epsilon l_B)^2 e^{i(\phi'-\phi)} (u_1^+ u_2^- - u_2^+ u_1^-
) 
\\ \nonumber
 &-& 2(v_1^+ v_2^- - v_2^+ v_1^-) 
 + i\sqrt{2} \epsilon l_B  \\ &\times& \nonumber
\left( e^{i\phi'} (v_1^+ u_2^- + u_2^+ v_1^-) +
e^{-i\phi} (u_1^+ v_2^- + v_2^+ u_1^-) \right),
\end{eqnarray}
where we use the shorthand notation
\begin{eqnarray*}
u_1^{\pm}& \equiv & D^{}_{(\epsilon l_B)^2/2-1} 
\left(\pm \sqrt{2}(-d/l_B+p_y l_B)\right ),
\\
v_1^{\pm}& \equiv & D^{}_{(\epsilon l_B)^2/2} \left
(\pm \sqrt{2}(-d/l_B+p_y l_B)\right ).
\end{eqnarray*}
The related symbols $u_2^\pm, v_2^\pm$ follow by letting $-d\to d$. 
The resulting transmission probability $T(\phi)=|t|^2$ 
is shown in Figure \ref{fig1} for several parameter values ($\epsilon,d$) 
outside the perfectly reflecting regime specified in Eq.~(\ref{perfecref}). 
For a typical value of $B_0=4$~T, the magnetic length is $l_B= 13$~nm, and 
$\epsilon l_B = 1$ corresponds to $E= 44$~meV.
Standard fabrication and doping techniques should thus be sufficient
to enter the perfect reflection regime.
On the other hand, for sufficiently high energy and/or narrow barriers, 
 the $\phi$-dependent transmission profile in Fig.~\ref{fig1}
should be observable.

We now turn to a discussion of 
a circularly symmetric {\sl magnetic quantum dot},
defined by a radially inhomogeneous field ${\bf B}= B(r) \hat e_z$.
It is convenient to use complex variables 
$z=x+iy$ (not to be confused with the $z$-direction) and $\bar z= x-iy$,
and the corresponding derivatives
$\partial=\frac12(\partial_x-i\partial_y)$ and
$\bar\partial= \frac12(\partial_x+i\partial_y)$,
where $r=\sqrt{\bar z z}$. 
Writing in a similar manner $A=A_x+iA_y$ and $\bar A=A_x-iA_y$,
the magnetic field is $B(r) = -i (\partial A - \bar \partial \bar A)$, 
where $(e/c) A = i \varphi(r)/\bar z$ and $(e/c) \bar A = - i \varphi(r) / z$.
This gauge expresses the vector potential in terms of the 
magnetic flux $\varphi(r)$ through a disc of radius $r$ in units of the 
flux quantum $hc/e$,
\begin{equation}\label{fluxdef}
\varphi(r)= \frac{e}{c} \int_0^r dr' r' B(r').
\end{equation} 
Next, we recall that in 2D the group of rotations is $SO(2)\sim U(1)$, 
whose generator is the orbital angular momentum  operator
$L= z\partial-\bar z \bar\partial$,
with eigenfunctions $\sim z^m$  (integer $m$). 
For an isotropic field $B(r)$, 
the operator $J=L+\sigma_z/2$ is conserved, i.e.~eigenstates of Eq.~(\ref{h0})
are classified by the half-integer eigenvalue $j=m\pm 1/2$ of $J$,
\begin{equation} \label{angular}
\left( \begin{array}{c} \psi_+\\ \psi_- \end{array}\right) =    
\frac{1}{\sqrt{2\pi}} 
\left( \begin{array}{c} \phi_m(r) (z/r)^{m}
\\ \chi_m(r) (z/r)^{m+1} \end{array}\right) .   
\end{equation}
The Dirac equation (\ref{h0}) then reduces to a pair of radial 1D
equations for $\phi_m(r)$ and $\chi_m(r)$ (where $r>0$ and $f'=df/dr$),
\begin{eqnarray}\label{cup}
\phi^\prime_m - \frac{m+\varphi(r)}{r} \phi_m &=& i\epsilon \chi_m, \\
\nonumber
\chi^\prime_m + \frac{m+1+\varphi(r)}{r} \chi_m &=& i\epsilon \phi_m,
\end{eqnarray}
implying a second-order equation
 for the upper component of Eq.~(\ref{angular}),
\begin{equation}\label{2nd}
 \phi_m^{\prime\prime} + \frac{1}{r} \phi_m^\prime
 + \left ( \epsilon^2- \frac{e}{c}B(r)- 
\frac{(m+\varphi(r))^2}{r^2} \right) \phi_m  = 0,
\end{equation} 
plus a similar equation for $\chi_m(r)$.  For $\epsilon\neq 0$,
$\chi_m$ directly follows from $\phi_m$ via Eq.~(\ref{cup}).

We now analyze a simple model for a magnetic quantum dot,
where $B=B_0$ outside a disk of radius $R$ and zero inside,
as previously considered for Schr\"odinger fermions in Ref.~\cite{peeters4}.
The flux (\ref{fluxdef}) is with $l_B=\sqrt{c/eB_0}$ given by
\begin{equation}\label{potdot}
\varphi(r)=\frac{r^2-R^2}{2 l_B^2} \theta(r-R).
\end{equation}
With normalization constant ${\cal N}_m$ and $m\leq 0$, 
the states 
\[
\psi_{m}^{\epsilon=0} = {\cal N}_m \left(\frac{r}{R}\right)^{\theta(r-R)R^2/2l_B^2} e^{-\varphi(r)/2} \left (
\begin{array}{c} 0 \\ (z/r^2)^m \end{array} \right)
\]
represent zero-energy solutions of Eq.~(\ref{h0}). 
The remaining eigenspectrum comes in pairs $\pm \epsilon$, and
we focus on the $\epsilon>0$ sector.
Up to an overall normalization factor,
Eq.~(\ref{2nd}) implies Bessel function solutions inside the dot, 
$\phi^<_m=J_m(\epsilon r)$ for $r<R$.
The general solution $\phi^>$ outside the dot ($r>R$) involves the
degenerate hypergeometric functions $\Phi$ and $\Psi$ 
\cite{gradsteyn}. With $\xi=r^2/2l_B^2$ and $\tilde m=m-\delta$, 
where $\delta= R^2/2l_B^2$ is the
missing flux through the dot, we obtain
\begin{equation} \label{outer}
\phi^>_m  =   \xi^{|\tilde m|/2} e^{-\xi/2 } \left(a_1 
\Phi(\alpha,1+|\tilde m|;\xi) + a_2 \Psi(\alpha,1+|\tilde m|;\xi) \right).
\end{equation}
Here $a_{1,2}$ are arbitrary complex coefficients, 
and energy is parameterized by 
\begin{equation}\label{alpha}
\alpha=1+ \tilde m\theta(\tilde m) -(\epsilon l_B)^2/2 .
\end{equation}
Continuity of $\psi(r)$ at $r=R$ now implies continuity 
of both $\phi_m(r)$ and $\phi^\prime_m(r)$, see Eq.~(\ref{cup}).
The resulting two matching conditions then determine the possible eigenstates.

\begin{figure}[t!]
\scalebox{0.35}{\includegraphics{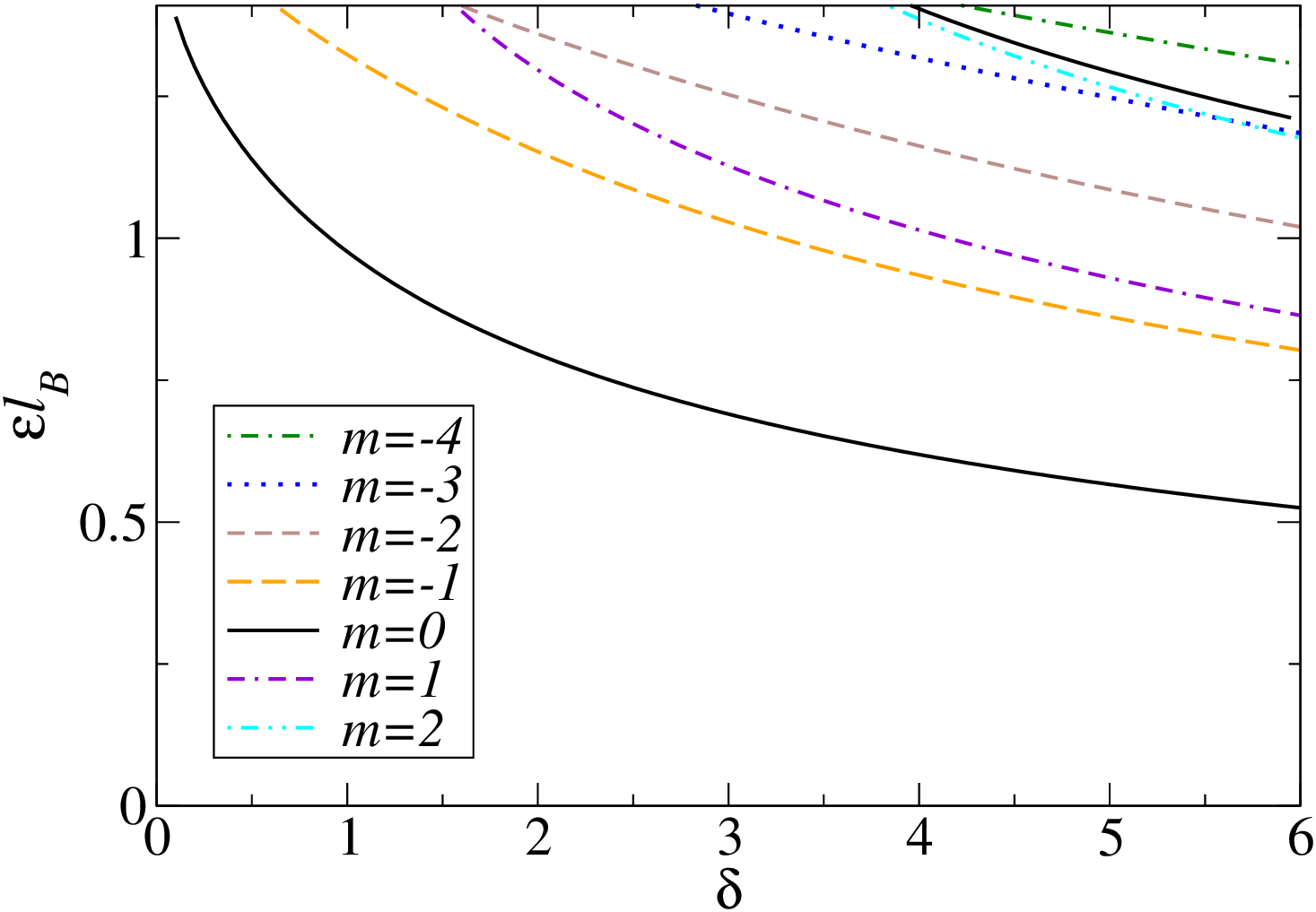}} 
\caption{ \label{fig2} (Color online)
Low-energy eigenenergies (labeled by $m$) 
for a disk-like magnetic quantum dot in graphene
versus missing flux $\delta=R^2/2 l_B^2$. 
}
\end{figure}

Note that the well-known relativistic Landau levels (for $R=0$) 
correspond to $\alpha=-n$ (with $n=0,1,2,\ldots$) \cite{semenoff},
where $\Phi$ and $\Psi$ reduce to Laguerre polynomials.
For finite $R$, the matching problem does {\sl not}\
admit solutions with $\alpha=-n$,
and we thus consider $\alpha\neq -n$.  
However, for $\alpha\neq -n$,  $\Phi$ has the asymptotic behavior
$\Phi\sim e^\xi$ at $\xi\to\infty$, i.e.~normalizability
of $\psi$ necessarily requires $a_1=0$ in Eq.~(\ref{outer}).  
One of the two conditions then fixes $a_2$, and the other 
determines the quantization condition on the energy,
\begin{eqnarray}\label{quantenerg}
&& 1-|\tilde m|\theta(-\tilde m)/\delta-\frac{\epsilon l_B}{\sqrt{2\delta}} 
\frac{J_{m+1}(\epsilon l_B \sqrt{2\delta})}{J_m(\epsilon l_B\sqrt{2\delta})}
 \\ \nonumber && =
 \frac{d}{d\xi}\ln \Psi(\alpha,1+|\tilde m|;\xi=\delta ).
\end{eqnarray}
The numerical solution of Eq.~(\ref{quantenerg})
is possible using standard root finding methods
(bracketing and bisection).
In Figure \ref{fig2}, we show the solutions to Eq.~(\ref{quantenerg})
with $\epsilon>0$ but below the lowest positive-energy bulk
Landau level located at $\epsilon l_B=\sqrt{2}$.
Within the shown $\delta$ range, for $m\neq 0$, there is 
at most one solution with $0<\epsilon l_B <\sqrt{2}$,
while for $m=0$, we obtain two such solutions for $\delta\agt 4$.
Depending on the missing flux $\delta\sim R^2 B_0$,
the energy levels of this 'Dirac dot' can be tuned almost at will.

To conclude, we have described a new way of confining
Dirac-Weyl quasiparticles in graphene.
We hope that our work will guide experimental efforts
to the development of mesoscopic structures
based on this novel material, and stimulate more theoretical
work on the effects of magnetic barriers on Dirac fermions.
 
We thank A. Altland, T. Heinzel, and W. H\"ausler for discussions.
This work was supported by the SFB TR 12 of the DFG
and by the ESF network INSTANS.


\begin{thebibliography}{99}
\bibitem{foot}
The Zeeman field can easily be included but does not
affect our conclusions.
\bibitem{foot1} The 
limit $B_0\to  \infty$ cannot be taken without spoiling consistency
of the single-valley approach.
\end{thebibliography}
\end{document}